\documentclass[aps,prb,twocolumn,showpacs,superscriptaddress,floatfix,nofootinbib]{revtex4-1}

\usepackage{graphicx}
\usepackage{amsmath}
\usepackage{epsfig}
\usepackage{helvet}
\usepackage{amssymb}

\newcommand{\be}{\begin{equation}}
\newcommand{\ee}{\end{equation}}
\newcommand{\bea}{\begin{eqnarray}}
\newcommand{\eea}{\end{eqnarray}}

\newcommand{\tr}{\mbox{tr}}
\newcommand{\bra}[1]{\mbox{$\langle #1 |$}}
\newcommand{\ket}[1]{\mbox{$| #1 \rangle$}}

\newcommand{\proj}[1]{\mbox{$|#1\rangle \!\langle #1 |$}}

\newcommand{\GS}{\mbox{\tiny GS}}

\def\tr{ \mbox{tr}}

\begin{document}

\title{Tensor network states and geometry}
\author{G. Evenbly$^{1}$, G. Vidal}
\affiliation{School of Mathematics and Physics, the University of
Queensland, Brisbane 4072, Australia}
\affiliation{Perimeter Institute for Theoretical Physics, Waterloo, Ontario, N2L 2Y5 Canada}
\date{\today}
 
\begin{abstract}
Tensor network states are used to approximate ground states of local Hamiltonians on a lattice in $D$ spatial dimensions. Different types of tensor network states can be seen to generate different geometries. Matrix product states (MPS) in $D=1$ dimensions, as well as projected entangled pair states (PEPS) in $D>1$ dimensions, reproduce the $D$-dimensional physical geometry of the lattice model; in contrast, the multi-scale entanglement renormalization ansatz (MERA) generates a $(D+1)$-dimensional holographic geometry. Here we focus on homogeneous tensor networks, where all the tensors in the network are copies of the same tensor, and argue that certain structural properties of the resulting many-body states are preconditioned by the geometry of the tensor network and are therefore largely independent of the choice of variational parameters. Indeed, the asymptotic decay of correlations in homogeneous MPS and MERA for $D=1$ systems is seen to be determined by the structure of geodesics in the physical and holographic geometries, respectively; whereas the asymptotic scaling of entanglement entropy is seen to always obey a simple boundary law -- that is, again in the relevant geometry. This geometrical interpretation offers a simple and unifying framework to understand the structural properties of, and helps clarify the relation between, different tensor network states. In addition, it has recently motivated the branching MERA, a generalization of the MERA capable of reproducing violations of the entropic boundary law in $D>1$ dimensions. 
\end{abstract}

\maketitle

\section{Introduction}
\label{sect:intro}

In recent years tensor network states 
\cite{Fannes92,Ostlund95,Rommer97,Perez-Garcia07, White92, White93, Schollwoeck05, Schollwoeck11, Vidal03, Vidal04, Daley04, White04, Shi06, Alba11, Vidal07, Vidal08, Evenbly09, Giovannetti08, Pfeifer09, Vidal10, Verstraete04, Sierra98, Nishino98, Nishio04, Murg07, Jordan08, Gu08, Jiang08, Xie09, Murg09, Tagliacozzo09, Murg10, Evenbly10, Evenbly10b, Aguado08, Cincio08, Evenbly09b, Koenig09, Evenbly10c, Corboz10, Kraus10, Pineda10, Corboz09, Barthel09, Shi09, Li10, Corboz10b, Pizorn10, Gu10, Corboz10c} 
have emerged as an important theoretical tool to investigate quantum many-body systems. They offer a novel conceptual framework to describe and classify the possible phases of matter\cite{Pollmann09,Chen11,Schuch10,Chen11b}. At the same time, as variational ans\"atze, tensor network states are the basis of numerical approaches to quantum many-body problems.

The simplest and best known tensor network state is the matrix product state (MPS)\cite{Fannes92,Ostlund95,Rommer97,Perez-Garcia07}. The MPS is at the core of the extraordinary success of White's density matrix renormalization group (DMRG) \cite{White92, White93, Schollwoeck05, Schollwoeck11}, which for almost twenty years has dominated numerical research in one dimensional lattice models, such as quantum spin chains, providing extremely accurate ground state properties. The MPS is also used to simulate dynamics with the time-evolving block decimation (TEBD)\cite{Vidal03, Vidal04, Daley04, White04} algorithm and variations thereof, often referred to as time-dependent DMRG. Other tensor network states for one-dimensional systems include the tree tensor network (TTN)\cite{Shi06, Alba11} and the multi-scale entanglement renormalization ansatz (MERA)\cite{Vidal07, Vidal08, Evenbly09, Giovannetti08, Pfeifer09, Vidal10}, with the later being particularly successful at describing ground states at quantum critical points.
  
Each of the above tensor network states for $D=1$ dimensional systems has a natural generalization in $D>1$ dimensions. The projected entangled-pair state (PEPS)\cite{Verstraete04, Sierra98, Nishino98, Nishio04, Murg07, Jordan08, Gu08, Jiang08, Xie09, Murg09} generalizes the MPS, whereas $D>1$ versions of TTN\cite{Tagliacozzo09, Murg10} and MERA \cite{Evenbly10, Evenbly10b, Aguado08, Cincio08, Evenbly09b, Koenig09, Evenbly10c} also exist. Among those generalizations, PEPS and MERA stand out for offering efficient representations of many-body wave functions, thus leading to scalable simulations in $D>1$ dimensions; and, importantly, for also being able to address systems that are beyond the reach of quantum Monte Carlo approaches due to the so-called sign problem, including frustrated spins\cite{Murg09,Evenbly10c} and interacting fermions\cite{Corboz10, Kraus10, Pineda10, Corboz09, Barthel09, Shi09, Li10, Corboz10b, Pizorn10, Gu10, Corboz10c}.

An attractive feature of tensor network states is that they are largely unbiased variational ans\"atze, in the sense that they are capable of representing many different types of ground states through a proper choice of variational parameters, as clearly witnessed by two decades of MPS explorations with DMRG\cite{White92,White93,Schollwoeck05,Schollwoeck11}. By increasing the bond dimension $\chi$ of the MPS\cite{Vidal03,Perez-Garcia07}, which governs the size of its tensors and therefore the number of variational parameters, more entanglement can be reproduced and a more accurate approximation to the ground state of a lattice model is obtained. 

As a matter of fact, an MPS can exactly reproduce any many-body state of the system provided that the bond dimension $\chi$ is large enough\cite{Vidal03, Perez-Garcia07}, although this will typically require a prohibitively large value of $\chi$, namely a value exponentially large in the system size -- leading to an inefficient representation. What makes the MPS interesting is that some moderately small value of $\chi$ is often already capable of accurately approximating the ground state of a local Hamiltonian in $D=1$ dimensions\cite{White92, White93, Schollwoeck05, Schollwoeck11}, as recently clarified by Hastings\cite{Hastings07,Hastings07b}. For instance, for most gapped Hamiltonians, an accurate MPS approximation is obtained already with some finite bond dimension $\chi$ that depends on $H$ but is essentially independent of the size of the system. For critical (and thus gapless) systems much of the same is true of the MERA\cite{Vidal07, Vidal08, Evenbly09, Giovannetti08, Pfeifer09}, which with a fixed bond dimension $\chi$ is capable of accurately reproducing large scale properties of the ground state, such as the asymptotic scaling of two-point correlators and of entanglement entropy.

Tensor network states in $D>1$ dimensions, such as PEPS\cite{Verstraete04, Sierra98, Nishino98, Nishio04, Murg07, Jordan08, Gu08, Jiang08, Xie09, Murg09} and MERA\cite{Evenbly10, Evenbly10b, Aguado08, Cincio08, Evenbly09b, Koenig09, Evenbly10c}, are also thought to be capable of accurately describing a large variety of ground states. This is supported both by growing numerical evidence and by the existence of analytical MERA\cite{Aguado08, Koenig09} and PEPS\cite{Buerschaper09,Gu09} constructions for some topologically ordered ground states in $D=2$ dimensions. However, a sharp increase of computational costs with bond dimension $\chi$ implies that, in practice, simulations in $D>1$ dimensions are restricted to small values of $\chi$. This restriction implies favouring low entangled states over more robustly entangled states. And as a result, and unless specific preventive measures are taken, tensor network states in $D>1$ dimensions may artificially favour local order over topological order, or may effectively open a gap in a critical system.

It is therefore important to understand how a finite bond dimension preconditions the properties of a given tensor network ansatz. The goal of our review paper is to collect together a number of previous results in this direction, which can be found scattered through the literature, and to present them under a simple and unifying framework. We consider homogeneous tensor network states with a finite bond dimension $\chi$ and look into those structural properties (in practice, scaling of correlators and of entanglement entropy) that follow from the way the tensors are connected into a network -- and are therefore independent of the choice of variational parameters. Those properties can be directly associated with properties of the (discrete) geometry reproduced by the network.

A main merit of this presentation is that it exposes, in very simple geometric terms, the main structural differences between the MPS and the scale invariant MERA in $D=1$ dimensions, while also explaining why these two tensor network states are a natural ansatz to represent ground states of gapped and gapless Hamiltonians, respectively. The situation turns out to be much less clear-cut in $D>1$ dimensions, where not only the scale invariant MERA but also PEPS can describe certain type of gapless ground states; in addition, both PEPS and MERA fail to describe another type of more robustly entangled, gapless ground states. A second merit of our presentation is that it allows us to express, again in simple geometric terms, the limitations experienced by PEPS and MERA in $D>1$ dimensions, while preparing the stage to announce how to overcome these limitations, namely by using tensor networks corresponding to more sophisticated geometries, as recently proposed in Ref. \onlinecite{Evenbly11}.
 
The rest of the paper is divided into sections as follows. Section \ref{sect:GS} briefly reviews the typical behaviour of correlations and entanglement entropy in ground states of local Hamiltonians. 
In Sect. \ref{sect:TN} we argue that a key difference between MPS/PEPS and MERA is in the geometry that these tensor network states describe. MPS and PEPS reproduce the $D$-dimensional \textit{physical} geometry of the lattice model, whereas the MERA describes a $(D+1)$-dimensional \textit{holographic} geometry, with the additional dimension corresponding to length scale. Then Sect. \ref{sect:correlation} points out that the decay of correlations in the MPS and the scale invariant MERA can be regarded as following from the structure of geodesics in the physical and holographic geometries, respectively. Similarly, Sect. \ref{sect:entropy} argues that the scaling of entanglement entropy in the MPS and the scale invariant MERA can be understood to follow from a simple boundary law in some appropriate region of the physical and holographic geometries.
Sect. \ref{sect:gapped} considers the holographic geometry of ground states of gapped systems and a related tensor network state, the finite range MERA, which in some sense interpolates between the MPS and the scale invariant MERA.

Finally, Sect. \ref{sect:discussion} argues that thinking about tensor network states in terms of geometry offers more than just an attractive way of presenting previously known results. As demonstrated by the recent proposal of the branching MERA, it can also stimulate further progress in the field. Specifically, the restriction experienced by PEPS and MERA to obeying the boundary law for entanglement entropy in $D>1$ dimensions is overcome by considering tensor network states that reproduce a $D+1$ holographic geometry with a more sophisticated structure, including branching in the length scale direction. 

\section{Ground states of local Hamiltonians}
\label{sect:GS}

Let $\mathcal{L}$ denote a lattice in $D$ spatial dimensions made of $N$ sites, where each site is described by a complex vector space $\mathbb{V}$, and let $H:\mathbb{V}^{\otimes N} \rightarrow \mathbb{V}^{\otimes N}$, with $H^{\dagger} = H$,  be a Hamiltonian that decomposes as a sum of terms, with each term involving only a few neighbouring sites.  
We refer to any such Hamiltonian, made of short-range interactions, as a \textit{local} Hamiltonian. The ground state $\ket{\Psi_{\GS}} \in \mathbb{V}^{\otimes N}$ of $H$ is the state that minimizes the expectation value $\bra{\Psi}H\ket{\Psi}$. Typically, the ground state $\ket{\Psi_{\GS}}$ of a given local Hamiltonian $H$ has a number of structural properties that are common to most ground states of local Hamiltonians. Here we consider two such structural properties: the behaviour of two-point correlation functions and the scaling of entanglement entropy.

\subsection{Correlations}
\label{sect:GS:corr}

Let us first consider the two-point correlator
\begin{equation}
	C(x_1,x_2) \equiv \langle P_{x_1} Q_{x_2} \rangle - \langle P_{x_1} \rangle \langle  Q_{x_2} \rangle,
\end{equation}
where $P_{x_1}$ and $Q_{x_2}$ denote two local operators acting on (a neighborhood of) sites $x_1,x_2\in \mathcal{L}$, and $\langle O  \rangle$ stand for the ground state expectation value $\bra{\Psi_{\GS}} O\ket{\Psi_{\GS}}$.
In most ground states of local Hamiltonians, 	$C(x_1,x_2)$ decays with the distance between positions $x_1$ and $x_2$. In the limit of large distances, this happens in one of two characteristic ways. When the Hamiltonian $H$ is gapped, correlations decay exponentially\cite{Hastings04},
\begin{equation}
	C(x_1,x_2) \approx e^{-|x_1-x_2|/\xi},
	\label{eq:Cgap}
\end{equation}
where $\xi\geq 0$ denotes a correlation length. Instead, when the Hamiltonian $H$ is gapless, correlations decay polynomially\cite{Sachdev99},
\begin{equation}
	C(x_1,x_2) \approx |x_1-x_2|^{-q},
	\label{eq:Ccrit}
\end{equation}
where $q\geq 0$ is some exponent. In this paper we refer to gapped systems also as non-critical systems, and to gapless systems as critical systems. A gapped/non-critical system that is close to a critical point may display two-point correlations that decay according to the more refined, combined form \cite{DiFrancesco97}
\begin{equation}
	C(x_1,x_2) \approx \frac{e^{-|x_1-x_2|/\xi}}{|x_1-x_2|^{q}},
	\label{eq:Cmixed}
\end{equation}
which is dominated by the polynomial decay for distances smaller than the correlation length, $|x_1-x_2| \ll \xi$, and reduces to the exponential decay of Eq. \ref{eq:Cgap} at larger distances, $|x_1-x_2| \gg \xi$.

\subsection{Entanglement entropy}
\label{sect:GS:entropy}

The amount of entanglement between a region $A$ of the lattice and the rest of the system can be measured by the entanglement entropy
\begin{equation}
	S(A) \equiv -\tr \left(\rho_A \log_2 \rho_A\right),
\end{equation}
where $\rho_A$ is the reduced density matrix for region $A$, obtained from the ground state $\ket{\Psi_{\GS}}$ by tracing out the rest of the system, denoted $B$,
\begin{equation}
	\rho_A \equiv \tr_{B} \proj{\Psi_{\GS}}.
	\label{eq:rhoA}
\end{equation}

In $D$ spatial dimensions, most ground states of local Hamiltonians obey a boundary law (often also referred to as "area law") for entanglement entropy\cite{Srednicki93, Latorre04, Plenio05, Bravyi06, Eisert06, Hastings07c, Masanes09, Eisert10}, in the sense that the entanglement entropy of a hyperblock $A$ of $L^{D}$ sites scales as the size $|\partial A|$ of its boundary $\partial A$,
\begin{equation}
	S(A) \approx |\partial A| \approx L^{D-1},~~~~~~\mbox{(boundary law)}
	\label{eq:boundaryLaw}
\end{equation}
instead of scaling as the size $|A| = L^{D}$ of the bulk of the block $A$. However, there are also ground states that display logarithmic corrections to the above boundary law \cite{Holzhey94, Callan94, Fiola94, Vidal03b, Wolf06, Gioev06, Li06, Barthel06, Swingle10, Swingle09b, Motrunich07, Senthil08, Liu09},
\begin{equation}
	S(A) \approx L^{D-1}\log_2 (L).
	\label{eq:LogD}
\end{equation}

Specifically, in $D=1$ dimensions, gapped systems obey a boundary law, which in this case means that the entanglement entropy saturates to a constant $S_0$ as a function of the block size $L$
\begin{equation}
	S(A) \leq S_0.
	\label{eq:1Dgap}
\end{equation}
Instead, critical systems display a logarithmic correction to the boundary law,  
\begin{equation}
	S(A) \approx \frac{c}{3} \log_2 (L),
\label{eq:1Dcrit}
\end{equation}
where $c$ is the central charge of the corresponding CFT. Near criticality, the entropy grows with $L$ as in Eq. \ref{eq:1Dcrit} until the saturation constant $S_0$ is reached, with the later scaling with the correlation length $\xi$ as
\begin{equation}
	S_0 \approx \frac{c}{3} \log_2 (\xi).
\label{eq:1Dmixed}
\end{equation}
 
The scaling of Eqs. \ref{eq:1Dgap}-\ref{eq:1Dmixed} were first presented in Ref. \onlinecite{Vidal03b} for specific quantum spin chains and are consistent with previous entropy calculations in quantum field theory by Holzhey et al. \cite{Holzhey94}, Callan and Wilczek \cite{Callan94} and Fiola et al. \cite{Fiola94}. Subsequently, Jin and Korepin\cite{Jin04} formalized the quantum spin chain results, whereas Cardy and Calabrese\cite{Calabrese04, Calabrese05} formalised and generalized the quantum field theory calculations.

In $D>1$, the relation between the scaling of entanglement entropy and the existence of a gap in $H$ is less clear-cut. For instance, the study of possible scalings of entanglement entropy in the ground state of systems of free fermions shows that the boundary law is obeyed by gapped systems, as expected, but also for a class of critical systems (namely, systems with a Fermi surface of dimension $\Gamma$ smaller than $D-1$), whereas a second class of critical systems (with a Fermi surface of dimension $\Gamma = D-1$) display logarithmic multiplicative corrections to the boundary law\cite{Wolf06, Gioev06, Li06, Barthel06}, see Table \ref{table:fermions}. Such logarithmic corrections are also believed to be present in other gapless systems in $D>1$ dimensions, such as Fermi Liquids and spin Bose metals\cite{Swingle10, Swingle09b, Motrunich07, Senthil08, Liu09}.

\begin{table}[!htb]
\begin{tabular}{|lcccc|}
\hline
          & $~\textrm{gapped}$ $\; \; \; \; \; \; \; $ & $\Gamma=0$ $\; \; \; \; \; \; \; $ & $\Gamma=1$  $\; \; \; \; \; \; \;$  & $\Gamma=2$  $\; \; \; \; \; \; \; $ \\ 
\hline
   D=1    & $S_0$ &   $\log_2 (L)$ &   -               & -                  \\
   D=2    &  $L$           &   $L$        &   $L \log_2 (L)$      & -              \\
   D=3    &  $L^2$         &   $L^2$      &   $L^2$             & $L^2 \log_2 (L)$     \\
\hline
\end{tabular}
\caption{Scaling of entanglement entropy $S(A)$ of a block $A$ made of $L^D$ sites for the ground state of free fermion models on a $D$-dimensional lattice and with a $\Gamma$-dimensional Fermi surface.}
\label{table:fermions}
\end{table}

We emphasize that the above simplified characterizations of correlations and entanglement entropy touch only on those aspects that are needed for subsequent analysis, and ignore a number of other important results. For instance, there are additive corrections to the boundary law for entanglement entropy in topological phases, known as topological entanglement entropy\cite{Kitaev06, Levin06}. Although both MERA\cite{Aguado08, Koenig09} and PEPS\cite{Buerschaper09, Gu09} can describe topologically ordered systems and in particular account for the topological entanglement entropy, the latter is only an additive correction to the scaling of entanglement entropy and as such will play no role in our discussions.

\section{Geometry of tensor network states}
\label{sect:TN}

\begin{figure}[!tb]
\begin{center}
\includegraphics[width=8.5cm]{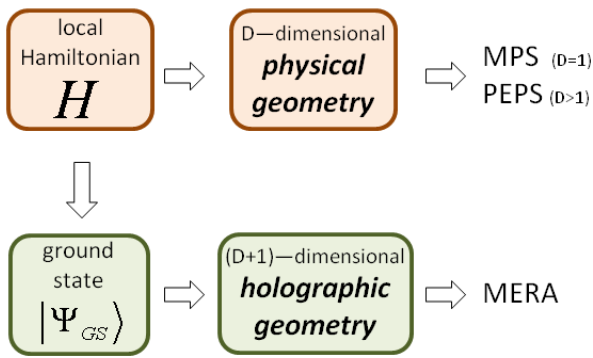}
\caption{ 
(Color online) 
A local Hamiltonian on a $D$-dimensional lattice defines a discrete $D$-dimensional geometry. To each ground state we can attach a $(D+1)$-dimensional geometry, where the additional dimension corresponds to length scale.
} \label{fig:geometries}
\end{center} 
\end{figure}

A tensor network state expresses the wave-function of a lattice system as a collection of tensors (i.e. multi-dimensional arrays of complex coefficients) that are connected according to a network pattern. We refer to the abundant literature for further details\cite{Fannes92,Ostlund95,Rommer97,Perez-Garcia07, White92, White93, Schollwoeck05, Schollwoeck11, Vidal03, Vidal04, Daley04, White04, Shi06, Alba11, Vidal07, Vidal08, Evenbly09, Giovannetti08, Pfeifer09, Vidal10, Verstraete04, Sierra98, Nishino98, Nishio04, Murg07, Jordan08, Gu08, Jiang08, Xie09, Murg09, Tagliacozzo09, Murg10, Evenbly10, Evenbly10b, Aguado08, Cincio08, Evenbly09b, Koenig09, Evenbly10c, Corboz10, Kraus10, Pineda10, Corboz09, Barthel09, Shi09, Li10, Corboz10b, Pizorn10, Gu10, Corboz10c}. 
In this paper we are concerned with the use of a tensor network state as an efficient, approximate representation of the ground state $\ket{\Psi_{\GS}}$ of a local Hamiltonian $H$ on a $D$-dimensional lattice. For concreteness, we consider square (or hypercubic) lattices, although most considerations can be easily generalised to other types of lattices. 

There are two \textit{geometries} that are relevant to the problem of representing a ground state, and most of the existing tensor network states can be broadly classified according to which of these two geometries their networks reproduce, see Fig. \ref{fig:geometries}. 

\begin{figure}[!tb]
\begin{center}
\includegraphics[width=7cm]{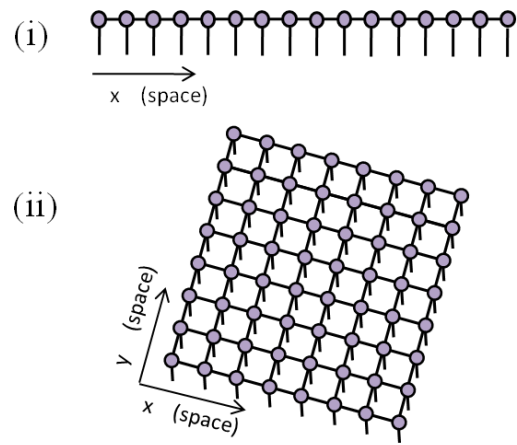}
\caption{ 
(Color online) 
(i) Matrix product state (MPS) for the ground state of a local Hamiltonian $H$ in a one-dimensional lattice. The tensors are connected according to a one-dimensional array, in correspondence with the one-dimensional physical geometry dictated by the interactions in $H$. (ii) Projected entangled pair state (PEPS) for the ground state of a two-dimensional lattice. The tensors are connected into a network that reproduces the two-dimensional physical geometry.
} \label{fig:MPS-PEPS}
\end{center} 
\end{figure}

\subsection{Physical geometry}
\label{sect:TN:physical}

First of all, there is the geometry generated by the pattern of interactions in $H$, which we refer to as \textit{physical geometry}. In a $D$-dimensional lattice $\mathcal{L}$, a short-ranged Hamiltonian $H$ connects neighbouring sites of $\mathcal{L}$. That is, two sites are close to each other in the physical geometry if and only if they are also close in the lattice $\mathcal{L}$. Therefore the physical geometry is also $D$-dimensional and essentially equivalent to the lattice $\mathcal{L}$ itself.

By definition, the physical geometry only depends on the pattern of interactions in Hamiltonian $H$ and is insensitive to any further details, such as whether $H$ is gapped or gapless. In particular, the physical geometry is also largely independent of specific structural properties of its ground state $\ket{\Psi_{\GS}}$, e.g. decay of correlations or scaling of entanglement entropy.

An important class of tensor network states consist in collections of tensors connected into a network that reproduces the physical geometry. For instance, a MPS reproduces the one-dimensional physical geometry of a spin chain, whereas PEPS reproduces the $D$-dimensional physical geometry in lattice models in $D>1$ spatial dimensions, see Fig. \ref{fig:MPS-PEPS}. If we regard the lattice model as a discretization of continuous space, then an infinite MPS describes a discretized version of the line, and an infinite PEPS in $D>1$ dimensions describes a discretized version of a $D$-dimensional hyperplane.

\begin{figure}[!tb]
\begin{center}
\includegraphics[width=7cm]{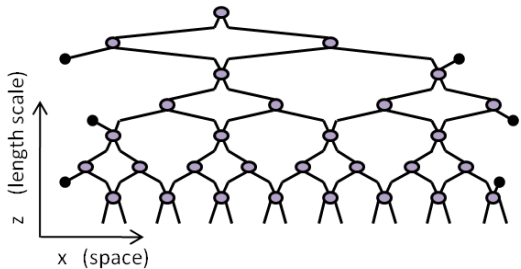}
\caption{ 
(Color online)
Multi-scale entanglement renormalization ansatz (MERA) for the ground state of a local Hamiltonian $H$ in a one-dimensional lattice. The tensors form a two-dimensional \textit{holographic geometry}. The horizontal direction reproduces the spatial dimension of the lattice model, whereas the vertical direction corresponds to the different length scales that are relevant to describing the structure of entanglement in the ground state of the system. More generally, the MERA for a system in $D$ dimensions spans a holographic geometry in $D+1$ dimensions.
} \label{fig:MERA}
\end{center} 
\end{figure}

\subsection{Holographic geometry}
\label{sect:TN:holographic}

A second geometry is given by the pattern of entanglement in the ground state $\ket{\Psi_{\GS}}$ of $H$. This pattern is naturally described by incorporating an additional dimension to the $D$-dimensional physical geometry. This additional dimension is associated to length scale (equivalently, energy scale), in the spirit of the holographic principle\cite{Maldacena98,Gubser98,Witten98,Swingle09}. Here we refer to the $(D+1)$-dimensional geometry generated in this way by the entanglement in the ground state $\ket{\Psi_{\GS}}$ as the \textit{holographic geometry} of $\ket{\Psi_{\GS}}$, and use the scale parameter $z$,
\begin{equation}
	z \equiv \log_2 \lambda,
\end{equation}
where $\lambda$ is a length scale, to label it. The physical geometry corresponds to setting $z=0$.
 
In the MERA, tensors are connected so as to reproduce the holographic geometry. For instance, the MERA for a one-dimensional system, Fig. \ref{fig:MERA}, spans two dimensions, thus describing a discrete, two-dimensional holographic geometry, with tensors labeled by a space coordinate $x$ and the scale parameter $z$. More generally, the MERA for the ground state of a $D$-dimensional system reproduces a discrete, $(D+1)$-dimensional holographic geometry. The MERA can be regarded as defining a real space renormalization group transformation, where the scale parameter $z$ labels coarse-grained lattices $\mathcal{L}^{(z)}$ that offer an effective description of the system at length scale $\lambda = 2^{z}$. This additional dimension allows the MERA to store, using different parts of the network, properties of the ground state corresponding to different length scales $\lambda$. 

In contrast with the physical geometry, which is only sensitive to the pattern of interactions in the local Hamiltonian $H$, the holographic geometry depends also on certain structural properties of the ground state, such as the existence of a finite correlation length $\xi$. In order to emphasize the differences between the physical and holographic geometries, here and in the next two sections we consider the holographic geometry of a critical, scale invariant ground state, in which the correlation length $\xi$ diverges and all length scales are equivalent. Correspondingly, in these sections we will restrict our considerations to the scale invariant MERA. Only later, in Sect. \ref{sect:gapped}, we will also address the case of a gapped Hamiltonian, where the correlation length $\xi$ is finite, and will consider a version of the MERA adequate to that situation.

\begin{figure}[!tb]
\begin{center}
\includegraphics[width=8.5cm]{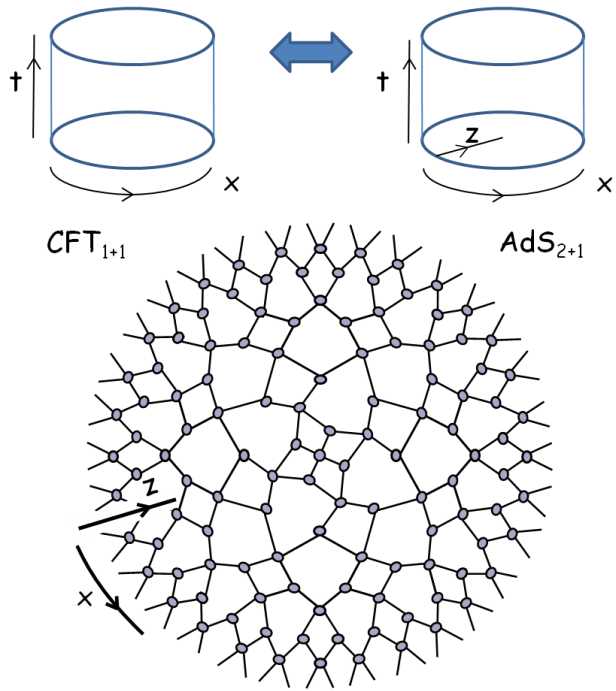}
\caption{ 
(Color online)
As pointed out by Swingle \cite{Swingle09}, the scale invariant MERA for the ground state of a quantum spin chain can be interpreted as a discrete realization of the AdS/CFT correspondence. The ground state of the one-dimensional lattice model corresponds to a discrete version of the vaccuum of a CFT$_{1+1}$, whereas the MERA spans a two dimensional geometry that corresponds to a discrete version of a time slice of AdS$_{2+1}$. The Figure shows a MERA similar to that of Fig. \ref{fig:MERA}, but from another perspective, with the scale parameter $z$ as a radial coordinate.
} \label{fig:AdSCFT}
\end{center} 
\end{figure}

The connection between MERA and the holographic principle was first explored by Swingle\cite{Swingle09}. Specifically, the scale invariant MERA\cite{Vidal07, Vidal08} used to describe the ground state of a quantum spin chain at criticality can be understood as a discrete realization of the AdS/CFT correspondence. Indeed, the (scale invariant) ground state of the critical chain is a discrete version of the vacuum of a $1+1$ conformal field theory (CFT), whereas the scale invariant MERA can be regarded as defining a discrete version of a $2+1$ anti de Sitter (AdS) space, see Fig. \ref{fig:AdSCFT}. [Notice that since we are describing time-independent ground states, the time direction is not captured by the tensor network and is irrelevant in the present discussion]. 

\subsection{Homogeneous tensor networks}

Several structural properties of the states that can be represented by the above tensor networks are pre-determined by the choice of geometry they reproduce. The next two sections review the behaviour of correlators and entanglement entropy in MPS, PEPS and MERA, and relate them to properties of the appropriate geometry. For simplicity, we mostly consider homogeneous tensor networks, in which all the tensors are copies of a single tensor (in the scale invariant MERA, two different tensors are necessary).

\begin{figure}[!tb]
\begin{center}
\includegraphics[width=8.5cm]{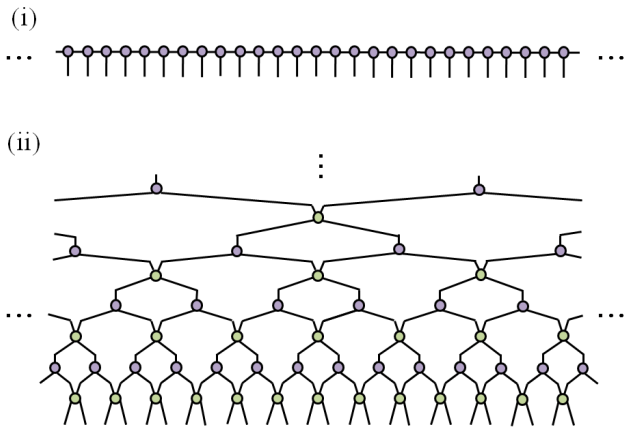}
\caption{ 
(Color online)
Homogeneous tensor network states for the ground state in an infinite lattice in $D=1$ spacial dimensions. (i) A homogeneous MPS is characterized by a single tensor that is repeated infinitely many times throughout the tensor network. (ii) A homogeneous scale invariant MERA is characterized by two tensors, a disentangler and an isometry, repeated throughout the tensor network, which consists of infinitely many layers.
} \label{fig:homogeneous}
\end{center} 
\end{figure}

We call \textit{generic} a property that is typically observed in a homogeneous tensor network where the coefficients of the tensor have been chosen randomly. In the following we consider \textit{generic} properties of \textit{homogeneous} MPS, PEPS and MERA with a finite bond dimension $\chi$. We consider states of an infinite lattice $\mathcal{L}$, see Fig. \ref{fig:homogeneous}, and focus mostly on the \textit{asymptotic} behaviour of such properties, namely in the decay of correlations at large distances and scaling of entanglement entropy for large blocks of sites.

\section{Correlations and geodesics}
\label{sect:correlation}

The asymptotic decay of correlations has long been known to be exponential in an MPS \cite{Fannes92, Ostlund95, Rommer97} and polynomial in the scale invariant MERA \cite{Vidal08, Giovannetti08, Pfeifer09}. In this section we point out that such behaviour is dictated by the structure of geodesics in the geometry attached to each of these tensor network states. For an MPS, the later is a rather straightfoward statement; for the MERA, it was first noted by Swingle\cite{Swingle09}. 
 
\begin{figure}[!tb]
\begin{center}
\includegraphics[width=7cm]{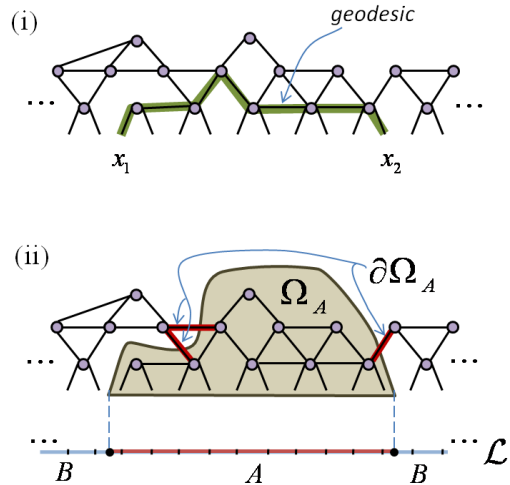}
\caption{ 
(Color online)
(i) Two sites $x_1$ and $x_2$ of the lattice $\mathcal{L}$ are connected through several paths within the tensor network. The geodesic corresponds to the shortest of such paths, where the length of a path is measured e.g. by the number of tensors in the path. In the example, the shortest path between sites $x_1$ and $x_2$ contains 6 tensors, and therefore the length of the geodesic is $D(x_1,x_2) = 6$.
Eq. \ref{eq:corrTN} relates the length of geodesics with the assymptotic decay of correlations in the tensor network (assuming that correlations are predominantly carried by the tensors in the geodesic). 
(ii) Region $\Omega_A$ of the tensor network that contains (the indices corresponding to) region $A$ of the lattice $\mathcal{L}$. The boundary $\partial \Omega_A$ of region $\Omega_A$ consists of the set of indices connecting $\Omega_A$ with the rest of the tensor network. The number $n(A)$ of such indices is interpreted as measure of the size $|\partial \Omega A|$ of the boundary $\partial \Omega_A$. In the example, $n(A) = |\partial \Omega_A|=3$. An upper bound to the entanglement entropy of a region $A$ of the lattice $\mathcal{L}$ is given in terms of $n(A)$, Eq. \ref{eq:upperBound}.
} \label{fig:geodesicTN}
\end{center} 
\end{figure}

\subsection{Geodesics within a tensor network}

Given a tensor network state for the state $\ket{\Psi}$ of a lattice $\mathcal{L}$, and two sites of $\mathcal{L}$ at positions $x_1$ and $x_2$, we can define a notion of distance between these two sites \textit{within the tensor network} as follows. First we notice that the two sites are connected by paths within the tensor network, where each path consists of a list of tensors and links/indices connecting the tensors. To any such path, we then associate a length, as given by the number of tensors (or links) in the path. Then the distance $D(x_1,x_2)$ between these sites is defined as the length of the shortest path connecting them, see Fig. \ref{fig:geodesicTN}(i). 

Let $C(x_1,x_2)$ denote a correlation function between positions $x_1$ and $x_2$. It turns out that for both the MPS and the scale invariant MERA, the decay of correlations can be expressed in terms of the distance $D(x_1,x_2)$ within the tensor network,
\begin{equation}
	C(x_1,x_2) \approx e^{-\alpha D(x_1,x_2)},
\label{eq:corrTN}
\end{equation}
for some positive constant $\alpha$.
This expression assumes that the correlations between the two sites are mostly carried through the tensors/links in the geodesic path connecting them. It originates in the fact that for both the MPS ($D=1$ dimensions) and the scale invariant MERA (in any dimensions), the correlator $C(x_1,x_2)$ can be obtained by evaluating an expression with the (approximate) form 
\begin{equation}
	C(x_1,x_2) \approx {\vec{v}_L}^{\dagger} \cdot (T)^{D(x_1,x_2)} \cdot \vec{v}_R, 
\label{eq:corrTNexplain}
\end{equation}
that is, a scalar product involving two vectors ${\vec{v}_L}$ and $\vec{v}_R$ and the $D(x_1,x_2)$-th power of some transfer matrix $T$. The eigenvalues of matrix $T$ give rise to the possible correlation lengths $\xi$ in Eq. \ref{eq:Cgap} for the MPS and the possible power laws $p$ in Eq. \ref{eq:Ccrit} for the scale invariant MERA. Instead of reproducing the original derivation of this result for MPS\cite{Fannes92, Ostlund95, Rommer97} and MERA\cite{Vidal08, Giovannetti08, Pfeifer09}, here we will focus on the geometrical interpretation of Eq. \ref{eq:corrTN} in terms of the structure of geodesics.

\subsection{Correlations in the MPS}
\label{sect:correlation:MPS}

The MPS reproduces the physical geometry of a lattice $\mathcal{L}$ in $D=1$ dimensions, and therefore the induced \textit{physical distance},
\begin{equation}
	D_{\text{\tiny phys}}(x_1,x_2) \approx |x_1-x_2|,
	\label{eq:Dphys}
\end{equation}
is simply proportional to the number of lattice sites between positions $x_1$ and $x_2$, see Fig. \ref{fig:geodesic}(i). Replacing the physical distance in Eq. \ref{eq:corrTN} leads to the following asymptotic expression for the correlators of the MPS,
\begin{equation}
	C_{\text{\tiny MPS}}(x_1,x_2) \approx e^{-\alpha D_{\text{\tiny phys}}(x_1,x_2)} \approx e^{-|x_1-x_2|/\xi},
\label{eq:CMPS}
\end{equation}
for some correlation length $\xi>0$, which indeed reproduces the exponential decay of correlations characteristic of gapped systems, see Eq. \ref{eq:Cgap}.

\subsection{Correlations in the scale invariant MERA}
\label{sect:correlation:MERA}

\begin{figure}[!tb]
\begin{center}
\includegraphics[width=8.5cm]{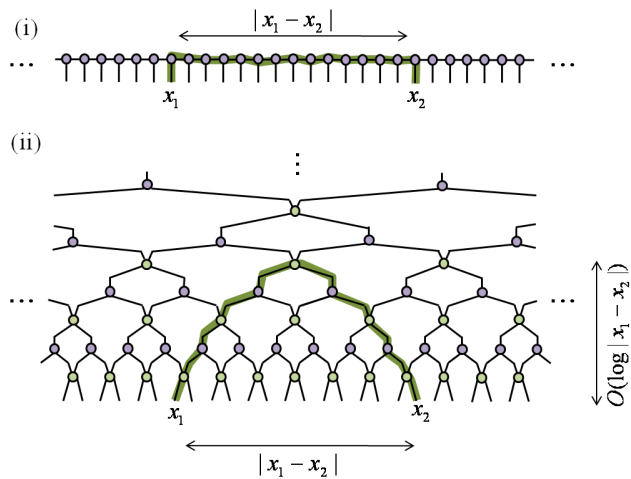}
\caption{ 
(Color online)
(i) In an MPS, two spins at positions $x_1$ and $x_2$ are connected by a path containing $|x_1-x_2|$ tensors. (ii) In a MERA, the same two spins are connected by a path that only has $O(\log_2(|x_1 - x_2|))$ tensors, in correspondence with geodesics in AdS space.
} \label{fig:geodesic}
\end{center} 
\end{figure}

In the scale invariant MERA, two sites at positions $x_1$ and $x_2$ of the lattice $\mathcal{L}$ are connected by a geodesic path of length $O(\log_2(x_1-x_2))$, see Fig. \ref{fig:geodesic}(ii), giving rise to the holographic distance
\begin{equation}
	D_{\text{\tiny{hol}}}(x_1,x_2) \approx \log_2(|x_1-x_2|), 
\label{eq:Dhol}
\end{equation}
which is consistent with the structure of geodesics in AdS space\cite{Swingle09}. Replacing this holographic distance in Eq. \ref{eq:corrTN} leads to the following asymptotic expression for the correlators of the scale invariant MERA, 
\begin{eqnarray}
	C_{\text{\tiny{MERA}}}(x_1,x_2) &\approx& e^{-\alpha D_{\text{\tiny hol}}(x_1,x_2)} \\ &\approx& e^{-q\log_2(|x_1-x_2|)} = |x_1-x_2|^{-q},
	\label{eq:CMERA}
\end{eqnarray}
for some exponent $q\geq 0$, which reproduces the polynomial decay of correlators characteristic of critical systems, Eq. \ref{eq:Ccrit}.

\subsection{Correlations in $D>1$ dimensions}

In $D>1$ dimensions, parts of the same analysis can be conducted again for the PEPS and the scale invariant MERA. The physical geometry reproduced by the PEPS induces a physical distance $D_{\text{\tiny{phys}}}(x_1,x_2)$ which, as in Eq. \ref{eq:Dphys}, is proportional to the distance within the lattice $\mathcal{L}$, whereas the scale invariant MERA leads to a holographic distance $D_{\text{\tiny{hol}}}(x_1,x_2)$ analogous to that of Eq. \ref{eq:Dhol}. It is also true that, for a generic choice of tensors in the PEPS and MERA, we again recover an asymptotic decay of correlation functions $C(x_1,x_2)$ that is exponential and polynomial in $|x_1-x_2|$, respectively, see Eqs. \ref{eq:CMPS} and \ref{eq:CMERA}.

However, we point out that for certain (non-generic) choices of variational parameters, PEPS can also display polynomial decay of correlations, as is the case e.g. of a PEPS built from a critical classical partition function\cite{Verstraete06}. Such (non-generic) behaviour is incompatible with the assumption implicit in Eq. \ref{eq:corrTN}, namely that correlations are mostly carried by the tensors/links included in the geodesic path connecting positions $x_1$ and $x_2$. In a PEPS with polynomially decaying correlation functions, correlations between two sites are instead obtained from a sum of contributions involving the many different paths connecting the two sites within the $D$-dimensional network, and not just from the geodesic paths. Therefore, Eq. \ref{eq:corrTN} does not hold for critical PEPS.

\begin{figure}[!tb]
\begin{center}
\includegraphics[width=6cm]{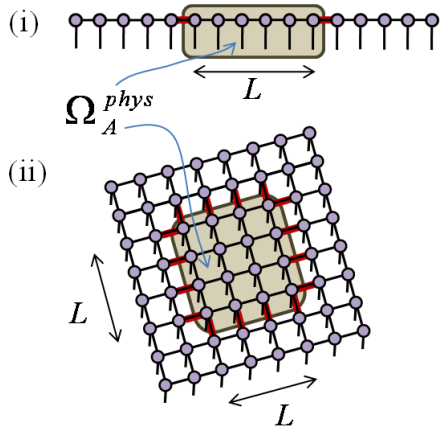}
\caption{ 
(Color online)
Upper bound for the entropy $S_L$ of the reduced density matrix $\rho_L$ of a region $A$ of linear size $L$: (i) In an MPS, the tensors describing a block $A$ of $L$ sites, that is region $\Omega_A^{\text{\tiny{phys}}}$, are connected with the rest of the tensor network by means of two (that is, a constant number of) bonds, $n(A)=2$. Therefore an MPS can at most reproduce a constant entanglement entropy $S_L \approx \mbox{const.}$, which corresponds to a (physical) boundary law. (ii)  In a PEPS for a two-dimensional system, the number $n(A)$ of bond indices connecting $\Omega_A^{\text{\tiny{phys}}}$ with the rest of the tensor network is proportional to the size of the boundary of the square region $A$, $n(A)\approx L$. Therefore the entropy scales at most as $S_{L} \approx L$, which again is a boundary law.
} \label{fig:entropyMPS}
\end{center} 
\end{figure}

\section{Entanglement entropy and boundary laws}
\label{sect:entropy}

The scaling of the entanglement entropy $S(A)$ of a region $A$ is well understood for a MPS\cite{Vidal03,Perez-Garcia07}, a PEPS\cite{Verstraete06} and a scale invariant MERA\cite{Vidal08pre}. In this section we review these results and re-express them as a simple boundary law for a related region $\Omega_A$ in the appropriate geometry.

\subsection{Entropy as the size of a boundary}

An upper bound for the entanglement entropy $S(A)$ in a tensor network state is obtained as follows. First we consider splitting the tensor network into two parts, $\Omega_A$
and $\Omega_B$, where $\Omega_A$ contains the open indices corresponding to all the sites in region $A$ and the other part $\Omega_B$ contains the open indices corresponding to region $B$, namely the rest of sites in lattice $\mathcal{L}$. Then we count the number of bond indices $n(A)$ that connect regions $\Omega_A$ and $\Omega_B$. Since each bond index can contribute at most $\log_2 (\chi)$ to the entropy of $\rho_A$, we obtain the upper bound\cite{entropyCount} (see Fig. \ref{fig:geodesicTN}(ii)),
\begin{equation}
	S(A) \leq n(A) \log_2 (\chi).
	\label{eq:upperBound}
\end{equation}
Among all such partitions, the one that minimizes $n(A)$ is the one that provides the tightest upper bond to $S(A)$. From now one, we use $\Omega_A$ and $\Omega_B$ to refer to this optimal partition, and $n(A)$ to denote the corresponding minimal number of bond indices.
The optimal upper bound of Eq. \ref{eq:upperBound} is saturated' for MPS, PEPS and MERA, in the sense that plenty of numerical evidence shows that for a generic choice of coefficients in a homogeneous tensor network, the entanglement entropy scales proportional to $n(A)$,
\begin{equation}
	S(A) \approx n(A).
\label{eq:SAnA}
\end{equation}

In our discrete geometries, we can think of $n(A)$ as a measure of the size $|\partial \Omega_A|$ of the boundary $\partial \Omega_A$ of the minimally connected region $\Omega_A$, and therefore interpret Eq. \ref{eq:SAnA} as stating that the entropy $S(A)$ is proportional to the size of the boundary of region $\Omega_A$,
\begin{equation}
	S(A) \approx |\partial\Omega_A|.
\label{eq:SAOA}
\end{equation}
This expression, equally valid for MPS, PEPS and MERA, allows us to always interpret the scaling of entanglement entropy as a simple boundary law in the appropriate geometry. The specific scaling of entanglement entropy for each tensor network state is then obtained by replacing $|\partial\Omega_A|$ in Eq. \ref{eq:SAOA} with its explicit dependence on the linear size $L$ of region $A$, as we do next.
 
\subsection{Entanglement entropy in MPS and PEPS}
\label{sect:entropy:MPS}

Consider a hypercubic region $A$ made of $L^D$ sites. The MPS and the PEPS reproduce the physical geometry. Therefore the hypercubic region $A$ of the lattice and the region $\Omega^{\text{\tiny{phys}}}_A$ of the tensor network that contains the open indices corresponding to sites in $A$ are essentially equivalent. In particular, the size of their boundaries is proportional, $|\partial \Omega^{\text{\tiny{phys}}}_A| \approx |\partial A|$. In other words, region $\Omega^{\text{\tiny{phys}}}_A$ is connected with the rest of the tensor network by a number of bond indices $n(A)$ proportional to the size $|\partial A|$ of the boundary $\partial A$ of the hypercubic region $A$ itself, see Fig. \ref{fig:entropyMPS}. Since $|\partial A| \approx L^{D-1}$, we obtain,
\begin{equation}
	n(A) \approx |\partial A| \approx L^{D-1},
\end{equation}
which implies, together with Eq. \ref{eq:SAnA}, that the entanglement entropy of the MPS\cite{Vidal03,Perez-Garcia07} and PEPS\cite{Verstraete06} scales with the linear size $L$ according to the boundary law of Eq. \ref{eq:boundaryLaw}, that is
\begin{eqnarray}
	S_{\text{\tiny MPS}}(A) &\approx& L^{D-1} \approx S_0 ~~~~~~(D=1) \label{eq:SMPS}\\
	S_{\text{\tiny PEPS}}(A) &\approx& L^{D-1} ~~~~~~~~~~~~~(D>1).
\end{eqnarray}
Here we will refer to such scaling of entanglement entropy as a \textit{physical boundary law} or simply \textit{boundary law}. 

\begin{figure}[!tb]
\begin{center}
\includegraphics[width=8.5cm]{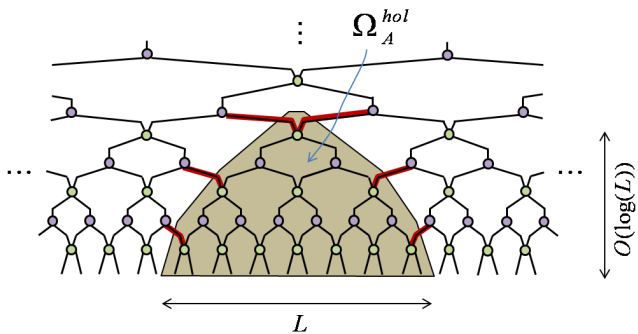}
\caption{ 
(Color online)
Upper bound for the entropy $S_L$ of the reduced density matrix $\rho_L$ of a region $A$ of $L$ contiguous sites. In a MERA for a one-dimensional system, the minimally connected region $\Omega_A^{\text{\tiny{hol}}}$ for region $A$ of the lattice is connected with the rest of the tensor network by a number $n(A)$ that grows logarithmically with the size of region $A$, $n(A) \approx \log (L)$. Therefore the entanglement entropy scales at most as $S_L \approx \log (L)$, which is a logarithmic violation of the (physical) boundary law. A more detailed analysis of the scaling is found in Fig. \ref{fig:OmegaA1D}.
} \label{fig:entropyMERA}
\end{center} 
\end{figure}
 
\subsection{Entanglement entropy in the scale invariant MERA} 
\label{sect:entropy:MERA}

Given a $D$-dimensional hypercubic region $A$ of lattice $\mathcal{L}$,  the minimally connected region $\Omega^{\text{\tiny{hol}}}_A$ in the scale invariant MERA is $(D+1)$-dimensional, see Fig. \ref{fig:entropyMERA}, with the additional dimension labelled by the scale parameter $z$. The number $n(A)$ of bond indices connecting $\Omega^{\text{\tiny{hol}}}_A$ with the rest of the tensor network is the sum of $T \approx \log_2 L$ different contributions $n_z(A)$, 
\begin{equation}
	n(A) \approx \sum_{z=0}^{T-1} n_z(A),
\label{eq:nA_MERA}
\end{equation}
where each contribution $n_z(A)$ corresponds to a different length scale $\lambda=2^z$, with $z\in \{0,1,\cdots, T-1\}$. As explained in Ref. \onlinecite{Vidal08pre}, the contribution $n_z(A)$ is proportional to the size of the boundary of a region $A_z$ obtained from region $A$ by means of $z$ coarse-graining steps, where each coarse-graining step divides the linear size of the region roughly by two, and where the size of the boundary of a region is measured by the number of boundary sites included in the region. Let us explicitly perform the sum in Eq. \ref{eq:nA_MERA}. It is useful to address $D=1$ and $D>1$ separately.

\begin{figure}[!tb]
\begin{center}
\includegraphics[width=6cm]{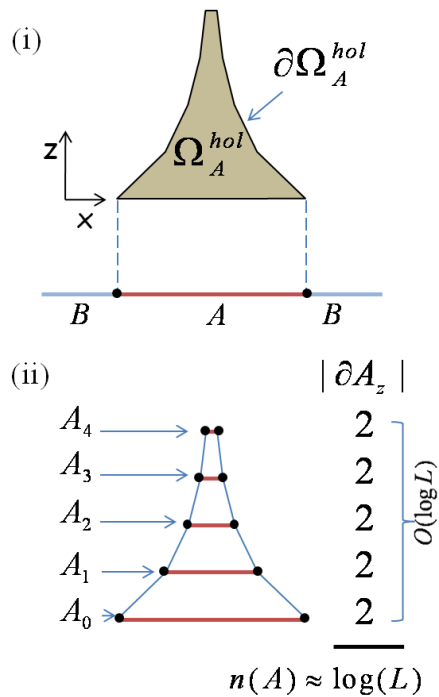}
\caption{ 
(Color online)
Scaling of entanglement entropy in the MERA in $D=1$ dimensions. (i) Caricature of region $A$ in the lattice $\mathcal{L}$ and of the corresponding region $\Omega_A^{\text{\tiny{hol}}}$ in the MERA (see also Fig. \ref{fig:entropyMERA}). (ii) The total number $n(A)$ of bond indices connecting $\Omega_A^{\text{\tiny{hol}}}$ with the rest of the tensor network is the result of $\log (L)$ identical contributions, each corresponding to a different length scale or value of $z$. Thus, $n(A) \approx \log_2(L)$. As a result, the entropy of region $A$ in the MERA for $D=1$ dimensions scales at most as $S(A) \approx \log_2(L)$, which is a logarithmic violation of the (physical) boundary law.
} \label{fig:OmegaA1D}
\end{center} 
\end{figure}

In $D=1$ dimensions, given a region $A$ made of $L$ sites, each region $A_z$ has a boundary $\partial A_z$ made of two sites, so that each contribution $n_z(A) \approx |\partial A_z| = 2$ is constant. Therefore $n(A)$, which is made of $T\approx \log_2(L)$ constant contributions (Fig. \ref{fig:OmegaA1D}), grows logarithmically with $L$,
\begin{equation}
	n(A) \approx 2T \approx \log (L). 
\label{eq:nLogL}
\end{equation}
Then Eqs. \ref{eq:SAnA} and \ref{eq:nLogL} imply that the entanglement entropy in the scale invariant MERA in $D=1$ dimensions grows as\cite{Vidal08pre}
\begin{equation}
	S_{\text{\tiny{MERA}}}(A) \approx \log (L)~~~~~~~~~~~~(D=1),
\label{eq:SMERA1D}
\end{equation}
which reproduces the scaling characteristic of ground states of quantum critical systems, see Eq. \ref{eq:1Dcrit}.

\begin{figure}[!tb]
\begin{center}
\includegraphics[width=6cm]{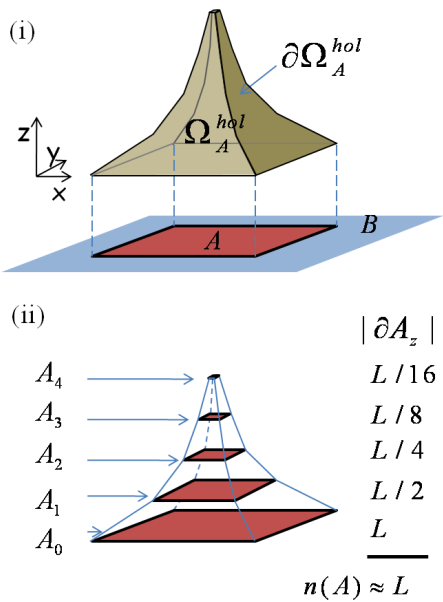}
\caption{ 
(Color online)
Scaling of entanglement entropy in the MERA in $D=2$ dimensions. (A similar analysis applies to $D>2$ dimensions). (i) Caricature of region $A$ in the lattice $\mathcal{L}$ and of the corresponding region $\Omega_A^{\text{\tiny{hol}}}$ in the MERA. (ii) The total number $n(A)$ of bond indices connecting $\Omega_A^{\text{\tiny{hol}}}$ with the rest of the tensor network is the result of $\log (L)$ contributions $n_z(A)$. Contribution $n_z(A)$ corresponds to length scale $\lambda = 2^z$ and is proportional to the size $|\partial A_z|\approx L/2^{z}$ of the boundary $\partial A_z$ of a coarse-grained region $A_z$. The sum of contributions is dominated by the smallest length scale, $z=0$, and is thus proportional to $L$. As a result, the entropy of region $A$ in the MERA for $D=2$ dimensions scales at most as $S(A) \approx L$, which is a (physical) boundary law.
} \label{fig:OmegaA2D}
\end{center} 
\end{figure}

Instead, in $D>1$ dimensions, each region $A_z$ is a hypercubic block of size $\approx L/2^{z}$ (Fig. \ref{fig:OmegaA2D}), and therefore the size $|\partial A_z|$ of its boundary $\partial A_z$ scales with $L$ as
\begin{equation}
	|\partial A_z| \approx \left(\frac{L}{2^{z}}\right)^{D-1}.
\end{equation}
Using again that $n_z(A)$ is proportional to $|\partial A_z|$ we find that now the contributions $n_z(A)$ to $n(A)$ in Eq. \ref{eq:nA_MERA} depend on $L$. Their sum leads to 
\begin{equation}
	n(A) \approx L^{D-1} \sum_{z=0}^{T} 2^{-z} \approx L^{D-1}.
	\label{eq:nLD1}
\end{equation}
Then Eqs. \ref{eq:SAnA} and \ref{eq:nLD1} imply that the entanglement entropy in the scale invariant MERA in $D>1$ dimensions grows as\cite{Vidal08pre}
\begin{equation}
	S_{\text{\tiny{MERA}}}(A) \approx L^{D-1}~~~~~~~~~~~~(D>1),
\label{eq:SMERA2D}
\end{equation} 
which reproduces the scaling characteristic of ground states in most gapped systems and some gapless systems in $D>1$ dimensions, see Eq. \ref{eq:boundaryLaw}.

Eq. \ref{eq:SMERA1D} corresponds to a logarithmic violation of the (physical) boundary law in $D=1$ dimensions, whereas Eq. \ref{eq:SMERA2D} corresponds to the (physical) boundary law. Here we reinterpret both Eq. \ref{eq:SMERA1D} and Eq. \ref{eq:SMERA2D} as a \textit{holographic boundary law}, that is, as a boundary law in the region $\Omega^{\text{\tiny hol}}_A$ of the holographic geometry. 

This geometric interpretation of the scaling of entanglement entropy in the scale invariant MERA is inspired by (and can be considered a lattice version of) the results of Ref. \onlinecite{Ryu06, Ryu06b, Nishioka09}, where the entanglement entropy of a CFT is computed using the holographic principle, by noticing that it scales as the size of the boundary of a region $\Omega^{\text{\tiny hol}}_A$ with minimal boundary. 
We emphasize, however, that while Refs. \onlinecite{Ryu06, Ryu06b, Nishioka09} discuss the scaling of entanglement entropy in the actual ground state of a physical theory, our present discussion only concerns the scaling of entanglement entropy in a variational ansatz (which we hope to be a good approximate representation of ground states).

One merit of this geometric interpretation is that it motivates a strategy to build tensor network states that violate the boundary law also in $D>1$, as presented in Ref. \onlinecite{Evenbly11} and discussed in Sect. \ref{sect:discussion}.

\section{Holographic geometry in gapped systems}
\label{sect:gapped}

The holographic geometry considered so far in Sects. \ref{sect:TN}-\ref{sect:entropy} corresponds to scale invariant, critical ground states, as described by the scale-invariant MERA. This particular scenario has been used there to emphasize the differences between physical and holographic geometries, which are most evident for critical systems. However, all ground states, whether corresponding to a critical system or a non-critical one, have a holographic geometry. For completeness, in this section we consider the holographic geometry of the ground states of gapped systems, which was first discussed by Swingle\cite{Swingle09}. These ground states can be represented by a finite range MERA\cite{Evenbly09} -- a MERA with a finite number of layers of tensors, where tensors in different layers are in principle allowed to be different. 

Since the finite range MERA is not a homogeneous tensor network, its structural properties do not only depend on the way the tensors are connected into a network -- different layers of the MERA may contribute differently to, say, correlations and entanglement entropy. However, even in this case geometrical considerations alone will already allow us to reproduce some of the key properties that differentiate the ground states of gapped systems from critical ones. In addition, near criticality, where the correlation length $\xi$ is much larger than the lattice spacing, we will recover aspects of the scaling of correlations and entanglement entropy as a function of $\xi$. Finally, in the opposite limit --namely when the correlation length $\xi$ is of the order of the lattice spacing-- we will see that the holographic geometry reduces to the physical geometry. Correspondingly, for ground states close to this limit, a finite range MERA representation becomes equivalent to an MPS/PEPS representation.

\subsection{Correlation length, finite range MERA and truncated holographic geometry}

Let us then consider the ground state \ket{\Psi_{\GS}} of a gapped Hamiltonian $H$ in $D$ dimensions, in which correlations decay exponentially with distance according to Eq. \ref{eq:Cgap} (or possibly Eq. \ref{eq:Cmixed}) and therefore have a characteristic length scale, the correlation length $\xi$. 

Notice that if we coarse-grain the lattice according to a scheme that maps a block of $2^D$ sites into one site, after one coarse-graining step the correlation length $\xi$ has shrunk by a factor two, $\xi \rightarrow \xi' = \xi/2$. By applying more coarse-graining steps the correlation length will shrink further. In particular, after
\begin{equation}
	z_\xi \equiv \log_2(\xi)
\end{equation}
coarse-graining steps the correlation length will become one (in units of separation between lattice sites), and a few additional coarse-graining steps, say a fixed number $\Delta z$ (independent of $\xi$), will render all two-point correlators negligible (i.e. smaller than some pre-determined, small constant). That is, starting with a ground state \ket{\Psi_{\GS}} with correlation length $\xi$, it takes 
\begin{equation}	
	z_0 \equiv z_{\xi} + \Delta z \approx O(\log_2(\xi))
\end{equation}
steps of coarse-graining to produce a state with negligible two-point correlators.

Here we will assume that after the $z_{0}$ steps of coarse-graining (according to an entanglement renormalization scheme\cite{Vidal07,Vidal10}) the original ground state $\ket{\Psi_{\GS}}$ of the system has been transformed into a state that can be well approximated by the product state
\begin{equation}
	\ket{\text{prod}} \equiv \ket{0}\otimes \ket{0} \otimes \cdots \otimes \ket{0},
	\label{eq:prod}
\end{equation}
namely a state with no correlations between different lattice sites. State $\ket{\text{prod}}$ describes the ground state at a fixed-point of the RG flow that corresponds to a gapped phase without topological order.

\begin{figure}[!tb]
\begin{center}
\includegraphics[width=8.5cm]{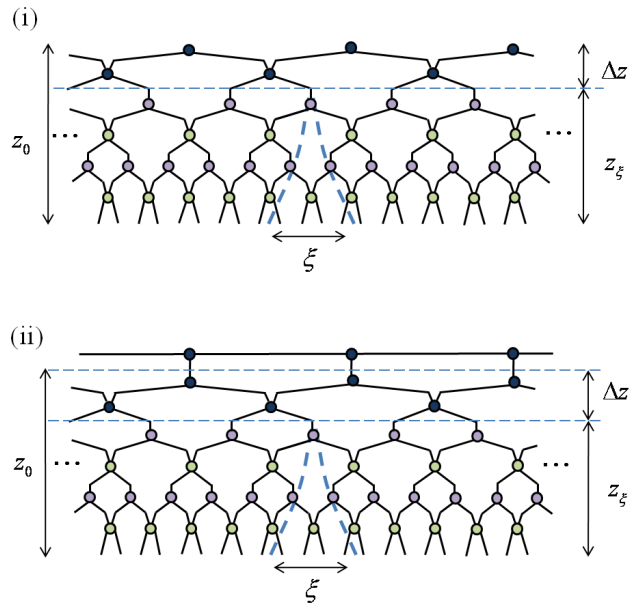}
\caption{ 
(Color online)
Finite range MERA for an infinite lattice in $D=1$ spatial dimensions. 
(i) $z_{\xi}\equiv \log_2 (\xi)$ layers of identical disentanglers and isometries obtained from the scale invariant MERA for the critical case are followed by some fixed number $\Delta z$ of non-homogeneous layers of tensors (where the tensors on different layers are allowed to be different). In this simple example, $z_{\xi} = 2$ and $\Delta z =1$, so that the total number of layers of tensors is $z_0 \equiv z_{\xi} + \Delta z = 3$. The finite range MERA represents a ground state $\ket{\Psi_{\GS}}$ of a gapped system that can be transformed into the unentangled state $\ket{\text{prod}}$ after three layers of coarse-graining.
(ii) The finite range MERA can be combined with another tensor network (e.g. MPS in the figure) in order to represent a ground state $\ket{\Psi_{GS}}$ of a gapped system that flows towards an entangled fixed point ground state $\ket{\Psi_{\mbox{\tiny f.p.}}}$ under coarse-graining transformations. The MPS at the top of the MERA can also be used to accurately describe the exponential decay of correlations that dominate the limit $|x_1-x_2|\gg \xi$ of Eq. \ref{eq:Cmixed}.
} \label{fig:frMERA}
\end{center} 
\end{figure}

In this case, the ground state $\ket{\Psi_{{\GS}}}$ of an infinite system can be represented by a \textit{finite range} MERA\cite{Evenbly09}, which is made of just a finite number $z_0$ of layers of disentanglers and isometries. The finite range MERA is not homogeneous, in that the tensors in different layers are allowed to be different, reflecting the fact that the properties of the ground state are now different at different length scales. Here we will consider a particular choice of finite range MERA, where the first $z_{\xi}$ layers of tensors correspond to those in the scale invariant MERA that describes the neighbouring critical point, and the remaining $\Delta z$ layers are chosen to minimize the ground state energy of the gapped Hamiltonian $H$. This choice is illustrated in Fig. \ref{fig:frMERA}(i) for a finite range MERA in $D=1$ dimensions.\cite{otherFiniteRangeMERA}

The holographic geometry attached to the ground state of a gapped system is still $(D+1)$-dimensional, but it is truncated in the RG direction, with the scale parameter $z$ restricted to values in the interval $[0,z_0]$. This truncation has an immediate effect on the possible decay of correlations and scaling of entanglement entropy that the tensor network can reproduce.

\subsection{Correlators}

\begin{figure}[!tb]
\begin{center}
\includegraphics[width=8.5cm]{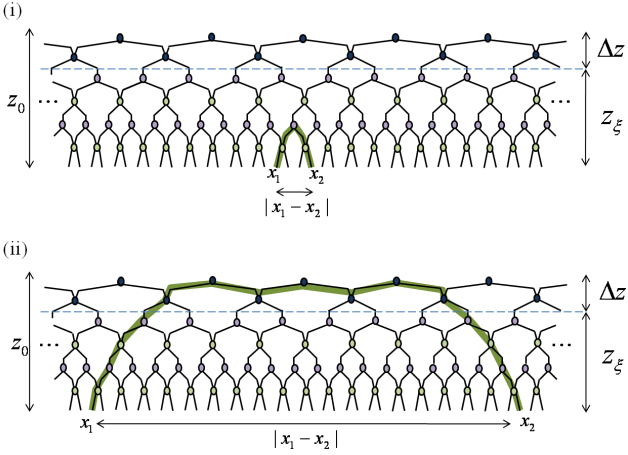}
\caption{ 
(Color online)
Geodesics in the finite range MERA. 
(i) When $|x_1-x_2|$ is smaller than the correlation length $\xi$, the geodesic connecting sites $x_1$ and $x_2$ within the tensor network is identical to the scale invariant case, and its length is therefore logarithmic in $|x_1-x_2|$.
(ii) When $|x_1-x_2|$ is larger than the correlation length $\xi$, the geodesic connecting the two sites sees the presence of the truncation and grows proportional to $|x_1-x_2|$. 
The structure of geodesics in the truncated holographic geometry is therefore compatible with the refined decay of correlations of Eq. \ref{eq:Cmixed}.  
} \label{fig:frMERAgeodesic}
\end{center} 
\end{figure}

Let us consider first a correlator between two sites $x_1$ and $x_2$ such that $|x_1-x_2|$ is smaller than the correlation length $\xi$, $|x_1-x_2| \ll \xi$. In this case the geodesic connecting the two sites within the truncated holographic geometry only runs through length scales $z$ smaller than $z_0$ and therefore is not affected by the existence of the truncation at $z=z_0$. As a result, the length of the geodesic is still logarithmic in $|x_1-x_2|$ as in a critical system, Eq. \ref{eq:Dhol},  see Fig. \ref{fig:frMERAgeodesic}(i). In addition, since the geodesic only runs through the homogeneous, scale-invariant region of the finite range MERA, $z\in [0,z_{\xi}]$, the two-point correlator is expected to decay polynomially as in a critical system, Eq. \ref{eq:Ccrit}. 

On the other hand, when $|x_1-x_2|$ is larger than the correlation length $\xi$, $|x_1-x_2|\gg \xi$, the length of the geodesic connecting the two sites within the tensor network grows proportional to $|x_1-x_2|$, see Fig. \ref{fig:frMERAgeodesic}(ii),
\begin{equation}
	D_{\text{\tiny hol}}(x_1,x_2) \approx |x_1-x_2|,~~~~ |x_1-x_2|\gg \xi ~~~~\text{(gapped $H$)}
\label{eq:CgapMERA}
\end{equation}
That is, for sufficiently large $|x_1-x_2|$, distances $D_{\text{\tiny hol}}(x_1,x_2)$ in the holographic geometry become proportional to distances $D_{\text{\tiny phys}}(x_1,x_2)$ in the physical geometry.

It would be tempting to say that, in this second regime, the structure of geodesics in the finite range MERA, Eq. \ref{eq:CgapMERA}, implies that correlations at large distances decay exponentially. While it is the case that the finite range MERA can approximate exponentially decaying correlations \cite{corrMERA}, these follow from the use of different tensors in top $\Delta z$ layers of the network, and cannot be interpreted in simple geometric terms. Nevertheless, what is clear from geometric arguments (together with some underlying transfer mechanism for the propagation of correlations) is that a truncated holographic geometry can no longer give rise to polynomially decaying correlations at long distances, $|x_1-x_2| \gg \xi$, since the length of geodesics is no longer logarithmic, Eq. \ref{eq:CgapMERA}.

In conclusion, we have argued that the truncated holographic geometry of the ground state of a gapped system gives rise to a modified structure of geodesics that is compatible with the decay of two-point correlators expressed in Eq. \ref{eq:Cmixed}, with polynomial decay for $|x_1-x_2| \ll \xi$ and exponential decay for $|x_1-x_2| \gg \xi$.

\begin{figure}[!tb]
\begin{center}
\includegraphics[width=8.5cm]{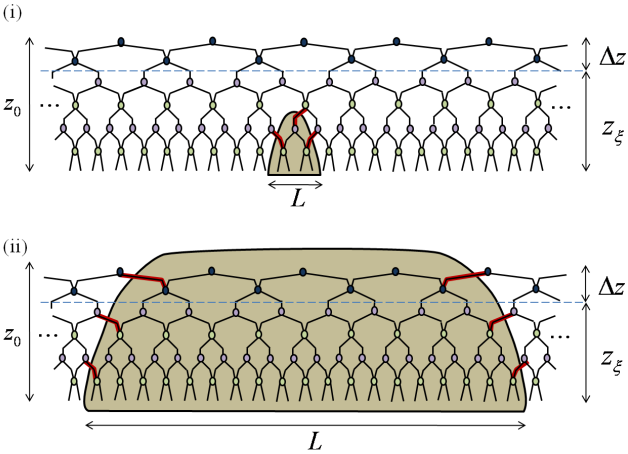}
\caption{ 
(Color online)
Minimally connected regions in the finite range MERA.
(i) When region $A$ is smaller than the correlation length $\xi$, $L\ll \xi$, the minimally connected region $\Omega_{A}^{\text{\tiny hol}}$ within the tensor network is identical to the scale invariant case, and the size $n(A) \equiv |\Omega_{A}^{\text{\tiny hol}}|$ of its boundary $\Omega_{A}^{\text{\tiny hol}}$ is therefore logarithmic in $L$.
(ii) When region $A$ is larger than the correlation length $\xi$, $L\gg \xi$, the size of the boundary $\Omega_{A}^{\text{\tiny hol}}$ saturates to a constant (as a function of $L$) that is proportional to the number $z_0$ of layers in the finite range MERA and thus grows logarithmically with $\xi$.
The structure of minimally connected regions in the truncated holography is therefore compatible with a saturation of the entropy, Eq. \ref{eq:1Dgap}, with a saturation value $S_0$ that scales as $\log_2(\xi)$, Eq. \ref{eq:1Dmixed}.
} \label{fig:frMERAentropy}
\end{center} 
\end{figure}

\subsection{Entanglement entropy}

In $D=1$ dimensions, the scaling of entanglement entropy in the finite range MERA (for gapped systems) is also different than in the scale invariant MERA (for critical systems), with the difference having a straightforward geometric interpretation. 

Let us consider Fig. \ref{fig:frMERAentropy}. If we first consider a region $A$ with length $L$ smaller than the correlation length $\xi$, $L \ll \xi$, then the minimally connected region $\Omega_A^{\text{\tiny hol}}$ in the tensor network only involves length scales $z$ smaller than $z_\xi$. As illustrated in Fig. \ref{fig:frMERAentropy}(i), in this case the size $n(A) \equiv |\partial \Omega_A^{\text{\tiny hol}}|$ of the boundary $\partial \Omega_A^{\text{\tiny hol}}$ scales as $\log_(L)$ as in the scale invariant MERA, Eq. \ref{eq:nLogL}, and the entanglement entropy grows as in a critical system, Eqs. \ref{eq:1Dcrit}.

However, when the size $L$ of region $A$ is larger than the correlation length $\xi$, $L \gg \xi$, the minimally connected region $\Omega^{\text{\tiny{hol}}}_A$ in the finite range MERA has a boundary $\partial \Omega^{\text{\tiny{hol}}}_A$ that saturates to a constant size $|\partial \Omega^{\text{\tiny{hol}}}_A|$, see Fig. \ref{fig:frMERAentropy}(ii), with 
\begin{equation}
	|\partial \Omega^{\text{\tiny{hol}}}_A| \equiv n(A) \approx z_\xi \equiv \log_2 (\xi) ~~~~(D=1,\text{ gapped $H$})
\end{equation}
That is, region $\Omega^{\text{\tiny{hol}}}_A$ is now connected with the rest of the tensor network through a number $n(A)$ of bond indices proportional to $z_{\xi} \equiv \log_2 (\xi)$, which is independent of $L$. This implies a constant upper bound for the entanglement entropy
\begin{equation}
	S_{\text{\tiny MERA}}(A) \approx L^{D-1} \approx S_0 ~~~~(D=1,\text{ gapped $H$})
\end{equation}
and therefore the finite range MERA obeys the boundary law of Eq. \ref{eq:1Dgap}.  
In addition, the entropy saturates to a constant $S_0\approx z_{\xi}$ that grows logarithmically with the correlation length $\xi$\cite{entMERA}, thus also producing a scaling compatible with Eq. \ref{eq:1Dmixed}.

\begin{figure}[!tb]
\begin{center}
\includegraphics[width=8.5cm]{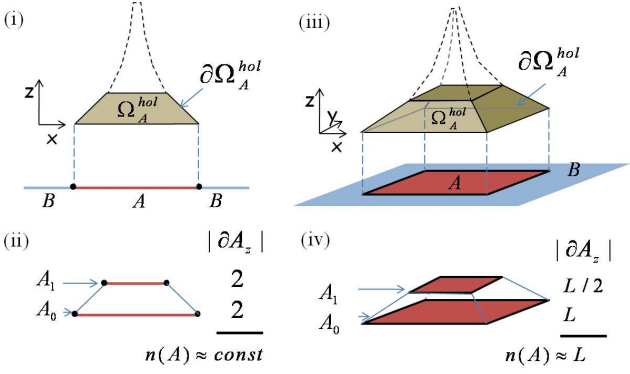}
\caption{ 
(Color online)
Scaling of entanglement entropy in the finite range MERA in $D=1$ and $D=2$ dimensions. (i) For $D=1$, region $\Omega_A^{\text{\tiny hol}}$ is a truncated version of that in Fig. \ref{fig:OmegaA1D}. (ii) Only length scales smaller than $\xi$ contribute to the total number $n(A)$ of bond indices connecting $\Omega_A^{\text{\tiny hol}}$ with the rest of the tensor network. This number is therefore upperbounded by a constant, which grows with the correlation length as $\log_2(xi)$. (iii) For $D=2$, the region $\Omega_A^{\text{\tiny hol}}$ is a truncated version of that in Fig. \ref{fig:OmegaA2D}. (iv) Again, only length scales smaller than $\xi$ contribute to the total number $n(A)$ of bond indices connecting $\Omega_A^{\text{\tiny hol}}$ with the rest of the tensor network. However, this does not change the linear dependence of $n(A)$ in $L$.
} \label{fig:OmegaAfinite}
\end{center} 
\end{figure}

Therefore we see that the structure of minimally connected regions in the truncated holographic geometry reproduces well the scaling of entanglement entropy in gapped systems, both for block lengths $L$ larger and smaller than the correlation length $\xi$.

In $D>1$ dimensions, the truncation of the holographic geometry to $z \leq z_0$ due to the presence of a finite correlation length $\xi$ does not alter the scaling of entanglement entropy, which is dominated by the $z=0$ contribution (see Fig. \ref{fig:OmegaAfinite}), and therefore the finite range MERA still obeys a boundary law,
\begin{equation}
	S_{\text{\tiny MERA}}(A) \approx |\partial \Omega_A^{\text{\tiny hol}}| \approx L^{D-1} ~~~~(D>1,\text{ gapped $H$})
\end{equation}

\subsection{Equivalence between holographic and physical geometries}

\begin{figure}[!tb]
\begin{center}
\includegraphics[width=8.5cm]{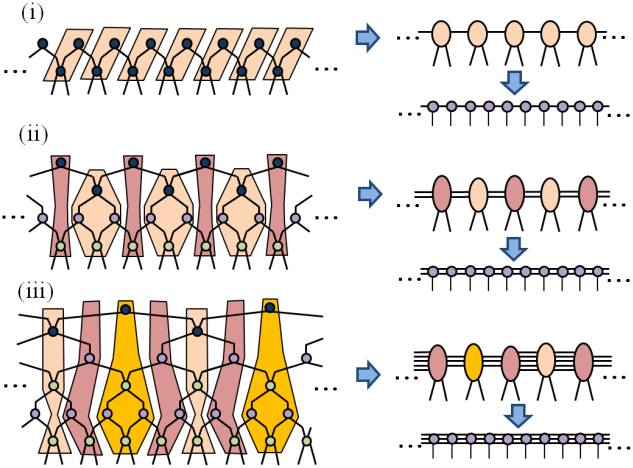}
\caption{ 
(Color online)
The finite correlation MERA can be converted into an MPS with a sufficiently large, but finite bond dimension $\chi_{\text{\tiny{MPS}}}$ as given by Eq. \ref{eq:MPSvMERA}. Specifically, each bond index of the MPS has to account for $O(\log_2 (\xi))$ bond indices of the MERA. When $\xi$ is small, the two-dimensional holographic geometry of $\ket{\Psi_{GS}}$ and the one-dimensional physical geometry of $H$ are essentially equivalent, and the ground state can be accurately described by either an MPS or a MERA. Figs. (i)-(iii) illustrate in diagrammatical notation how to convert a finite range MERA into a MPS, where the bond dimension of the MPS grows exponentially with the number of layers in the MERA.
} \label{fig:MERAtoMPS}
\end{center} 
\end{figure}

We have seen that for gapped systems, the holographic geometry is truncated at a value $z_0$ of the scale parameter $z$ corresponding (up to a constant) to $z_{\xi}\equiv\log_2(\xi)$, where $\xi$ is the correlation length. We have also seen that the presence of the truncation in the holographic geometry implies that the length $D_{\text{\tiny hol}}(x_1,x_2)$ of geodesics and the size $|\partial \Omega_{A}^{\text{\tiny hol}}|$ of the boundary of minimally connected regions in the holographic geometry scale asymptotically as in the physical geometry. As a matter of fact, when the correlation length $\xi$ is of the order of the lattice spacing, so that $z_0$ is a small number, it is no longer possible to distinguish between holographic and physical geometries at all. 

Correspondingly, as we discuss below, in $D=1$ dimensions the finite range MERA can be efficiently mapped into an MPS. This mapping if still possible for a large correlation length $\xi$, but the resulting MPS has a bond dimension $\chi$ that grows with $\xi$ and diverges at a critical point.

\subsection{From MERA to MPS}

As illustrated in Fig. \ref{fig:MERAtoMPS} for gapped systems in $D=1$ dimensions, a finite range MERA made of $z_0$ layers of tensors and with bond dimension $\chi_{\text{\tiny{MERA}}}$ can be re-expressed as an MPS with bond dimension $\chi_{\text{\tiny{MPS}}}$ given by\cite{MERAtoMPS}
\begin{equation}
	\chi_{\text{\tiny{MPS}}} \approx \left(\chi_{\text{\tiny{MERA}}}\right)^{z_0},~~~~~z_0\approx z_{\xi} \equiv \log_2(\xi).
	\label{eq:MPSvMERA}
\end{equation}
Therefore, when the correlation length $\xi$ is small, the holographic geometry is a narrow strip and the MERA can be re-expressed as an MPS with a small bond dimension $\chi_{\text{\tiny{MPS}}}$.

However, as one gets closer to a quantum critical point and the correlation length $\xi$ becomes larger, the holographic geometry becomes a strip with larger width $z_0$. The MERA can still be re-expressed as an MPS, but with a bond dimension $\chi_{\text{\tiny{MPS}}}$ that grows exponentially with the number $z_0$ of layers in the MERA, Eq. \ref{eq:MPSvMERA}. Since the computational cost of MPS algorithms scales as a power of $\chi_{\text{\tiny{MPS}}}$, that is exponentially with $z_0$, numerical simulations must be restricted to small values of $z_0$. 

Finally, at the critical point, where $\xi$ diverges, the holographic geometry extends indefinitely in the coarse-graining direction $z$, and a MERA with finite bond dimension $\chi_{\text{\tiny{MERA}}}$ can no longer be replaced by an MPS with finite bond dimension $\chi_{\text{\tiny{MPS}}}$.

In $D>1$ dimensions, a finite range MERA with a finite bond dimension $\chi_{\text{\tiny{MERA}}}$ can also be re-expressed as a PEPS of finite bond dimension $\chi_{\text{\tiny{PEPS}}}$. However, the bond dimension $\chi_{\text{\tiny{PEPS}}}$ does not grow significantly with $z_0$ and, as a matter of fact, even the scale invariant MERA in $D>1$ (with $z_0=\infty$) can be exactly represented by a PEPS with finite bond dimension $\chi_{\text{\tiny{PEPS}}}$, as recently shown in Ref. \onlinecite{Barthel10} (see also Ref. \onlinecite{PEPSinhomo}). 

\section{Discussion}
\label{sect:discussion}

In this manuscript we have reviewed a number of results concerning correlations and entanglement in tensor network states and presented them in a unified way by pointing out that they can be interpreted as geometric properties of some underlying discrete geometry.
Specifically, MPS and PEPS have been argued to describe a $D$-dimensional physical geometry dictated by the interactions of a local Hamiltonian $H$ in $D$-dimensions, whereas the MERA has been seen to describe a $(D+1)$-dimensional holographic geometry associated to the ground state $\ket{\Psi_{\GS}}$ of $H$. Our presentation clearly emphasizes the main structural differences between MPS/PEPS and MERA, and interprets the decay of correlations and the scaling of entanglement entropy in terms of geometric concepts such as geodesics and regions of minimal surface within the relevant geometry, as summarized in Eqs. \ref{eq:corrTN} and \ref{eq:SAOA}. The geometrical interpretation is also the natural language to connect the MERA with the holographic principle.

We conclude the present review with two brief discussions of related issues. The first is a practical warning for future tensor network practitioners. The second is a pointer to on-going developments that have been motivated by the geometric perspective described here.


\subsection{MPS for critical systems in $D=1$ dimensions; and for systems in $D=2$ dimensions.}

First, a word of caution on the use of geometric considerations to characterize tensor network states is in order. Our discussion has mostly focussed on the \textit{asymptotic} decay of correlations and \textit{asymptotic} scaling of entanglement entropy. In $D=1$ dimensions, this analysis pointed at the MPS as a natural representation for the ground state of gapped systems, and at the scale invariant MERA as a natural representation for the ground state of critical systems. However, this should not be understood as implying that an MPS cannot be used to study ground states of critical systems in $D=1$ dimensions, or even ground states in $D=2$ dimensional lattices. 

In a homogeneous MPS with finite bond dimension $\chi_{\text{\tiny MPS}}$, two-point correlators are indeed constrained to asymptotically decay exponentially, Eq. \ref{eq:CMPS}, whereas entanglement entropy must saturate, Eq. \ref{eq:SMPS}, thus reproducing the scaling of Eqs. \ref{eq:Cgap} and \ref{eq:1Dgap} characteristic of gapped systems. However, for some intermediate values of distance $|x_1-x_2|$ and size $L$, an MPS can still accurately approximate a polynomial decay $|x_1-x_2|^{-q}$ of correlations and a logarithmic growth $\log_2(L)$ of entanglement entropy. More specifically, a finite bond dimension $\chi_{\text{\tiny{MPS}}}$ in the MPS has been seen\cite{Tagliacozzo08, Nishino96, Pollmann09b} to introduce an artificial, finite correlation length $\xi_{\chi}$ (where $\xi_{\chi}$ depends on central charge $c$ of the CFT that describes the critical point under consideration) such that the correct scaling of correlators and entropy is reproduced for $|x_1-x_2|$ and $L$ smaller than $\xi_{\chi}$. 

A relatively mild scaling of computational costs with the bond dimension of the MPS, namely as $\chi_{\text{\tiny MPS}}$ to the third power, implies that very large bond dimensions (of the order of thousands) can be afforded with reasonably modest computational resources, leading to large values of the effective correlation length $\xi_{\chi}$. This, together with the use of finite size scaling techniques, make the MPS a very suitable tool to study critical ground states, which explains the success of DMRG also for critical systems\cite{White92, White93, Schollwoeck05, Schollwoeck11}. 

Similarly, an MPS may appear as an unlikely candidate to represent ground states of $D=2$ lattice models, since the only way it can afford reproducing the boundary law of entanglement entropy in $D=2$ dimensions, Eq. \ref{eq:boundaryLaw}, is through a bond dimension $\chi_{\text{\tiny MPS}}$ that grows exponentially in the linear size of the lattice. Once more, however, using an MPS with very large $\chi_{\text{\tiny MPS}}$ (which can again be afforded due to the relatively mild scaling of computational costs with $\chi_{\text{\tiny MPS}}$) and finite size scaling arguments, an MPS has been successfully used to study ground states of two-dimensional lattice models\cite{Liang94, White98, Xiang01, White07, Yan11}. 
 
\subsection{Beyond the entropic boundary law in $D>1$ dimensions.}

The present analysis has also reminded us of an important limitation of PEPS and MERA in $D>1$ dimensions. Recall that these tensor network states are constrained to obey a strict boundary law for entanglement entropy, Eq. \ref{eq:boundaryLaw}. However, there is an important class of gapless systems in $D>1$ dimensions whose ground states display a logarithmic violation of the boundary law, Eq. \ref{eq:LogD}. These systems include Fermi gases and liquids with a $(D-1)$-dimensional Fermi surface, as well as spin Bose-metals with an analogous Bose surface\cite{Swingle09b, Swingle10, Motrunich07, Senthil08, Liu09}. How may we go about using a tensor network state to represent such ground states?

In the case of PEPS, the boundary law cannot be easily overcome, since it is an intrinsic property of the physical geometry that the ansatz reproduces. Mimicking the previous discussion on the use of MPS to study critical systems, one could, perhaps, study ground states with a logarithmic violation of the boundary law with by a PEPS by considering finite systems and by suitably increasing the bond dimension $\chi_{\text{\tiny PEPS}}$ with the system size. Then finite size scaling techniques could be used to extrapolate finite size results to the thermodynamic limit. However, the cost of PEPS simulations grows as a much larger power of the bond dimension $\chi_{\text{\tiny PEPS}}$ than in the case of MPS, confining $\chi_{\text{\tiny PEPS}}$ to small values and seriously limiting the viability of this strategy.

\begin{figure}[!tb]
\begin{center}
\includegraphics[width=8.5cm]{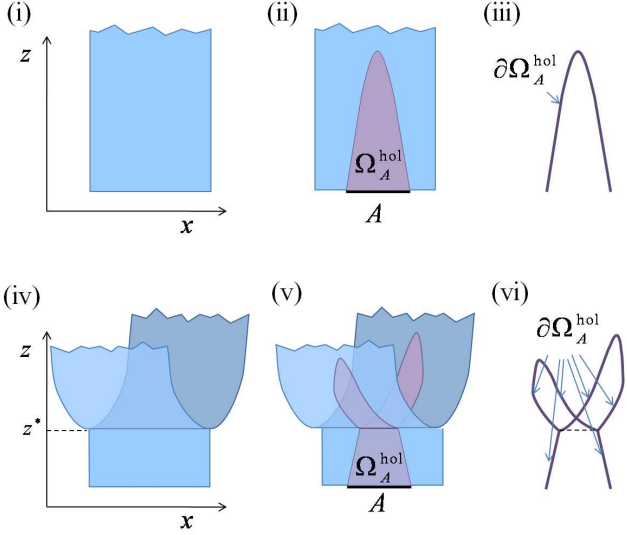}
\caption{ 
(Color online)
Holographic branching in a $D=1$ dimensional system. (i) Graphical representation of the holographic geometry in the absence of branching. Space is labelled by coordinate $x$ and whereas the scale parameter $z$ labels the different length scales $\lambda \equiv 2^z$ in the system. (ii) A region $A$ in the $D=1$ system defines a minimally connected region $\Gamma_A^{\text{\tiny hol}}$ in the the holographic geometry, as discussed in Sect. \ref{sect:entropy}. (iii) Boundary $\partial \Gamma_A^{\text{\tiny hol}}$ of the minimally connected region $\Gamma_A^{\text{\tiny hol}}$.
(iv) Graphical representation of the holographic geometry in the presence of branching at the value $z=z^{\star}$ of the scale parameter, corresponding to length scale $\lambda^{\star} \equiv 2^{z^{\star}}$. (v) The same region A of (ii) gives now rise to a minimally connected region $\Gamma_A^{\text{\tiny hol}}$ which is also affected by the branching (provided that the size $L$ of region $A$ is larger than the length scale $\lambda^{\star}$ at which branching occurs). (vi) As a result of the branching, the boundary $\partial \Gamma_A^{\text{\tiny hol}}$ of the minimally connected region $\Gamma_A^{\text{\tiny hol}}$ is larger. A larger boundary leads, together with Eq. \ref{eq:SAOA}, to a larger amount of entanglement entropy.
} \label{fig:BoundLawBranch}
\end{center} 
\end{figure}

Similar considerations apply to the MERA: a systematic increase of bond dimension $\chi_{\text{MERA}}$ as larger systems in $D>1$ dimensions are considered seems unviable, due to the sharp increase of computational costs with $\chi_{\text{\tiny MERA}}$. However, remember that in $D=1$ dimensions the logarithmic violation of the boundary law, Eq. \ref{eq:1Dcrit}, could be interpreted as a boundary law in the holographic geometry, Eqs. \ref{eq:SAOA} and \ref{eq:SMERA1D}. This strongly suggests a possible alternative. Indeed, it is natural to wonder whether, once more, the logarithmic violations of the boundary law in $D>1$ dimensions, Eq. \ref{eq:LogD}, can follow from a boundary law in a more elaborated, yet unknown, $(D+1)$-dimensional holographic geometry. A generalized MERA that would reproduce this holographic geometry would then automatically display a logarithmic violation of the boundary law.

\begin{figure}[!tb]
\begin{center}
\includegraphics[width=8cm]{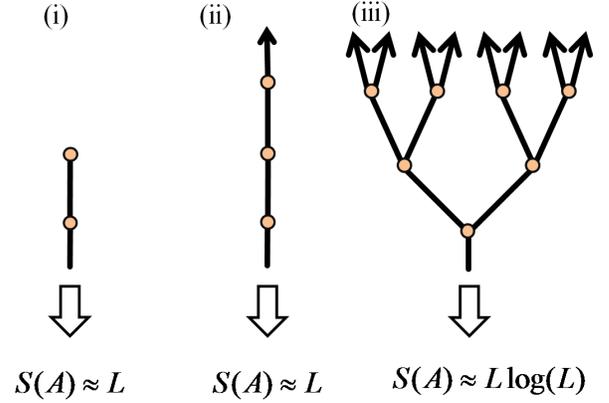}
\caption{ 
(Color online)
Schematic representation of different holographic geometries in $D=2$ dimensional systems in terms of the holographic tree introduced in Ref. \onlinecite{Evenbly11}. (i) The holographic geometry of a gapped system has a single branch with a finite extension in the $z$ direction, namely $z\in [0,z_0]$ where $z_0 = O(\log (\xi))$ as discussed in Sect. \ref{sect:gapped}. This geometry corresponds to a finite range MERA. (ii) The holographic geometry of a gapless system that obeys the entropic boundary law may also consist of a single branch, but this extends indefinitely in the $z$ direction, as in the scale-invariant MERA. 
(iii) Holographic geometry with an infinite number of branching points, capable of reproducing the logarithmic violation of the boundary law characteristic of the ground state of several gapless systems, including free fermions, see Table \ref{table:fermions}. Notice that the holographic geometry of (i) and (ii) allow us to distinguished between two types of ground states, corresponding to gapped and gapless systems, that obey the boundary law. In this sense, the holographic geometry can be used to issue a more refined classification of ground states according to their pattern of entanglement. 
} \label{fig:subsume}
\end{center} 
\end{figure}

\subsection{Holographic branching}
\label{sect:discussion:holBranch}

As recently discussed in Ref. \onlinecite{Evenbly11}, it turns out that, indeed, one can engineer holographic geometries such that Eq. \ref{eq:LogD}, as well as many other forms of scaling, can be understood to follow from a holographic boundary law. A key ingredient in these holographic geometries is the presence of branching, by means of which a single $(D+1)$-dimensional geometry associated to small length scales (high energies) becomes two independent $(D+1)$-dimensional geometries at large length scales (lower energies), see Fig. \ref{fig:BoundLawBranch} (i). Physically, holographic branching describes the decoupling of a single theory into two theories (or sets of degrees of freedom) that do not interact with each other at energy scales lower than some decoupling energy -- equivalently, at length scales larger than some decoupling length $\lambda$. Thus, at length scales smaller than $\lambda$ there is a single lattice model, whereas at length scales larger than $\lambda$, the lattice model breaks into two independent lattice models.

Fig. \ref{fig:BoundLawBranch} illustrates how the presence of holographic branching affects the amount of entanglement entropy in the ground state. The holographic region $\Omega_A$ associated with a physical region $A$ of linear size $L$ larger than $\lambda$ also branches into two pieces. As a result, the entropy $S(A)$ receives contributions from two pieces of the boundary $\partial \Omega_A$. As discussed in Ref. \onlinecite{Evenbly11}, it turns out that a sequence of holographic branchings occurring at different length scales, as represented by a branching tree (see Fig. \ref{fig:subsume}(iii) for an example) leads to a wide range of forms of scaling for the entanglement entropy $S(A)$ of a region $A$ in the original lattice, including Eq. \ref{eq:LogD}. The resulting tensor network state, the branching MERA, reproduces these model elaborated holographic geometries and has been shown to efficiently represent e.g. the ground state of a $D=2$ dimensional fermionic lattice model with a one-dimensional Fermi surface.

The study of holographic geometries with (possibly multiple) branching points opens up a number of exciting new possibilities, presently under consideration. On the one hand, it motivates a revision of the RG flow and its structure of fixed points. As we progress towards low energies, a single theory may branch into (perhaps infinitely many) other theories. In particular, new fixed points of this revised RG flow, including branching at all length scales, seem to include certain $D=2$ dimensional systems with a one-dimensional Fermi surface.

On the other hand, the holographic geometry also offers a new venue to characterize entanglement of ground states. The pattern of branching (as given by a holographic tree) of a ground state $\ket{\Psi_{\GS}}$, as well as the extent of each branch in the scale direction $z$, leads to a new classification of ground states that subsumes the one provided by considering only the scaling of the entanglement entropy $S(A)$ of a region $A$ of the lattice, see Fig. \ref{fig:subsume}. For instance, while gapped systems and some gapless systems in $D>1$ dimensions cannot be distinguished by the scaling of entanglement entropy (since they all obey the boundary law of Eq. \ref{eq:boundaryLaw}), their holographic geometry is clearly distinct.

This research was supported in part by the National Science Foundation under Grant No. NSF PHY05-51164. The authors also acknowledge support from the Australian Research Council under Grants FF0668731 and DP1092513.



\begin{thebibliography}{150}
\bibitem{MERAtoMPS} 
A related expression, namely $\chi_{\text{\tiny MPS}} \approx (\chi_{\text{\tiny MERA}})^{\log_2 N}$, was first quoted in the refereences of Ref. \onlinecite{Vidal08}, where it was erroneously written as $\chi_{\text{\tiny MPS}} \approx (\chi_{\text{\tiny MERA}}){\log_2 N}$.
\bibitem{entropyCount} The reduced density matrix $\rho_A$ of Eq. \ref{eq:rhoA} can be seen to have a rank $r(\rho_A)$ upper bounded by the total dimension of the $n(A)$ indices connecting parts $A$ and $B$ of the wavefunction, that is $r(\rho_A) \leq \chi^{n(A)}$, where $\chi$ is the dimension of each index. Then Eq. \ref{eq:upperBound} follows from the basic entropic inequality $\tr(\rho_A \log_2 \rho_A) \leq r(\rho_A)$, which is saturated when all the eigenvalues of $\rho_A$ are equal to $1/r(\rho_A)$.
\bibitem{otherFiniteRangeMERA}
The finite range MERA can also be used, in combination with another tensor network, to describe other gapped phases of matter, such as (possibly symmetry protected) topologically ordered phases, that are characterized by some entangled fixed-point ground state $\ket{\Phi_{\mbox{\tiny f.p.}}}$ instead of $\ket{\text{prod}}$. In this case, the finite range MERA is used to coarse-grain the ground state $\ket{\Psi_{\GS}}$ into $\ket{\Phi_{\mbox{\tiny f.p.}}}$, which is then represented by some other tensor network (e.g. an MPS, PEPS or scale invariant MERA). Fig. \ref{fig:frMERA}(ii) shows an example in D=1 dimensions.
\bibitem{corrMERA} Strictly speaking, the exponential decay of correlations in the finite range MERA does not accept a simple geometric interpretation. On the one hand, the ansatz is not homogenerous (that is, different tensors describe different length scales) and therefore the relation between the distance $D_{\text{hol}}(x_1,x_2)$ and a correlation function $C(x_1,x_2)$ is no longer be given by Eq. \ref{eq:corrTN}. On the other hand, in a finite range MERA all correlations $C(x_1,x_2)$ for distances $|x_1-x_2|$ larger than $O(2^{z_0})$ are zero, as a result of the isometric character of tensors in the MERA. This last limitation disappears if the tensors at the top layer of the MERA are not constrained to be isometric.
\bibitem{entMERA} In the MERA for a ground state near a quantum critical point, most of the entanglement entropy in a block of size $L\gg \xi$ comes from the bond indices corresponding from the first $z_0\equiv \log_2(\xi)$ layers of the finite range MERA, which make approximately equal contributions, so that the entropy saturates to a constant that depends on the correlation length $\xi$ as $S_{\text{\tiny{MERA}}}(A)\approx \log_2(\xi)$.
\bibitem{PEPSinhomo} The PEPS resulting from rewriting a scale invariant MERA in $D>1$ is significantly inhomogeneous (it requires infinitely many different tensors, one for each value of the scale parameter $z$) and its bond indices have to communicate correlations at all length scales. As a result, its practical use as a variational ansatz is not clear.
\end{thebibliography}
\end{document}